\documentclass[a4paper,11pt]{article}
\usepackage{pos}

\title{GRANDProto300: status, science case, and prospects}

\author*[a]{Simon Chiche} 
\onbehalf{for the GRAND Collaboration}

\affiliation[a]{Inter-University Institute For High Energies (IIHE), Université libre de Bruxelles (ULB), \\
Boulevard du Triomphe 2, 1050 Brussels, Belgium}

\emailAdd{simon.chiche@ulb.be}

\abstract{GRANDProto300, the mid-scale prototype of the GRAND experiment, is a planned radio array of 300 antennas over $200\, \rm km^{2}$ that will be deployed in the radio-quiet location of Xiao Dushan (China) by $\sim 2026$. The array will act as a test bench for the GRAND experiment and aim to achieve autonomous radio-detection and reconstructions of very inclined air showers in a large-scale array. GRANDProto300 will detect ultra-high-energy cosmic rays in the energy range $10^{16.5}-10^{18}\, \rm eV$ at a rate comparable to Auger. GRANDProto300 could also contain a ground particle array that would validate the performances of the radio detectors. We discuss the current status of the detector commissioning and the rich science case made possible by GRANDProto300, which covers the study of the Galactic-to-extragalactic transition, fast radio bursts and ultra-high-energy gamma-rays.}

\FullConference{10th International Workshop on Acoustic and Radio EeV Neutrino Detection Activities (ARENA2024)\\
11-14 June 2024\\
The Kavli Institute for Cosmological Physics, Chicago, IL, USA\\}

\begin{document}
\maketitle

\section{Introduction}

GRANDProto300 is the mid-scale prototype of the GRAND~\cite{GRAND, KoteraGRAND} experiment, designed to detect radio signals emitted by ultra-high-energy astroparticles interacting in the Earth's atmosphere. It will serve as a test bench for the GRAND experiment and probe the feasibility of radio-detection of astroparticles with a sparse large-scale radio array. Specifically, GRANDProto300 aims to demonstrate the feasibility of autonomous radio detection and the accurate reconstruction of inclined air showers in a radio-quiet environment. Its science case will range from the study of the Galactic-to-extragalactic transition to fast radio bursts and ultra-high-energy gamma rays. We present the current status of the experiment, its design, and expected performances.

\section{GRANDProto300 concept}

\subsection{Site location and radio array design}

 \begin{figure*}[tb]
\centering 
\includegraphics[width=0.38\columnwidth]{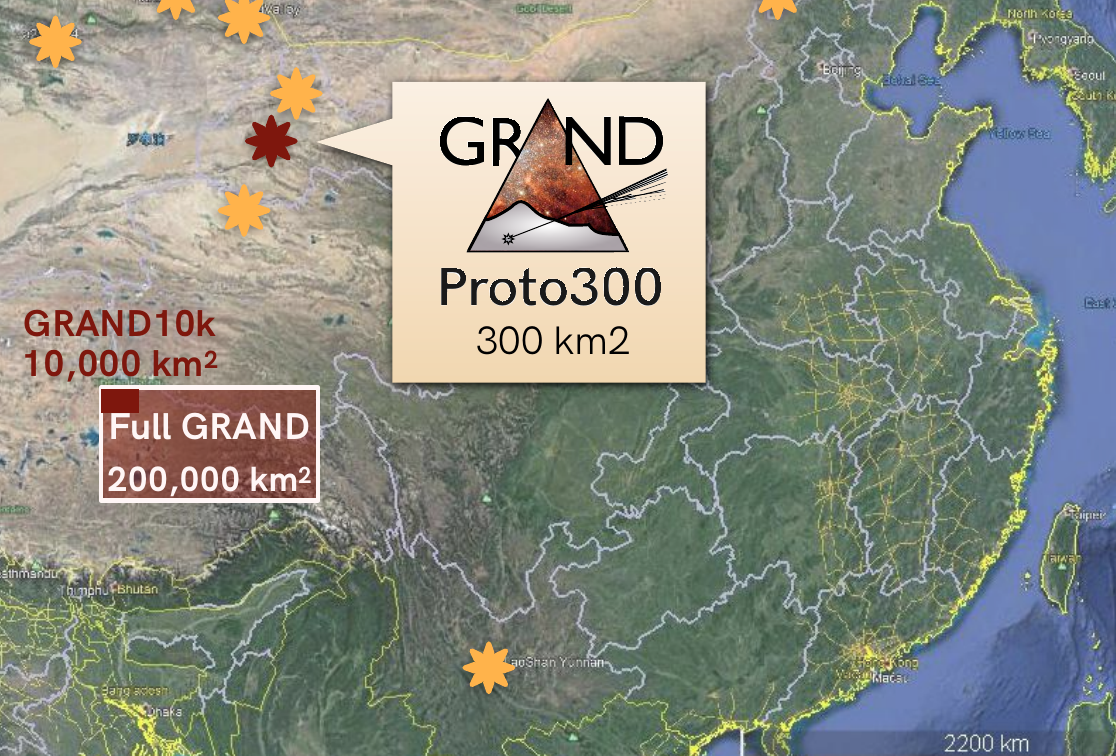}
\raisebox{-0.5cm}{\includegraphics[width=0.58\columnwidth]{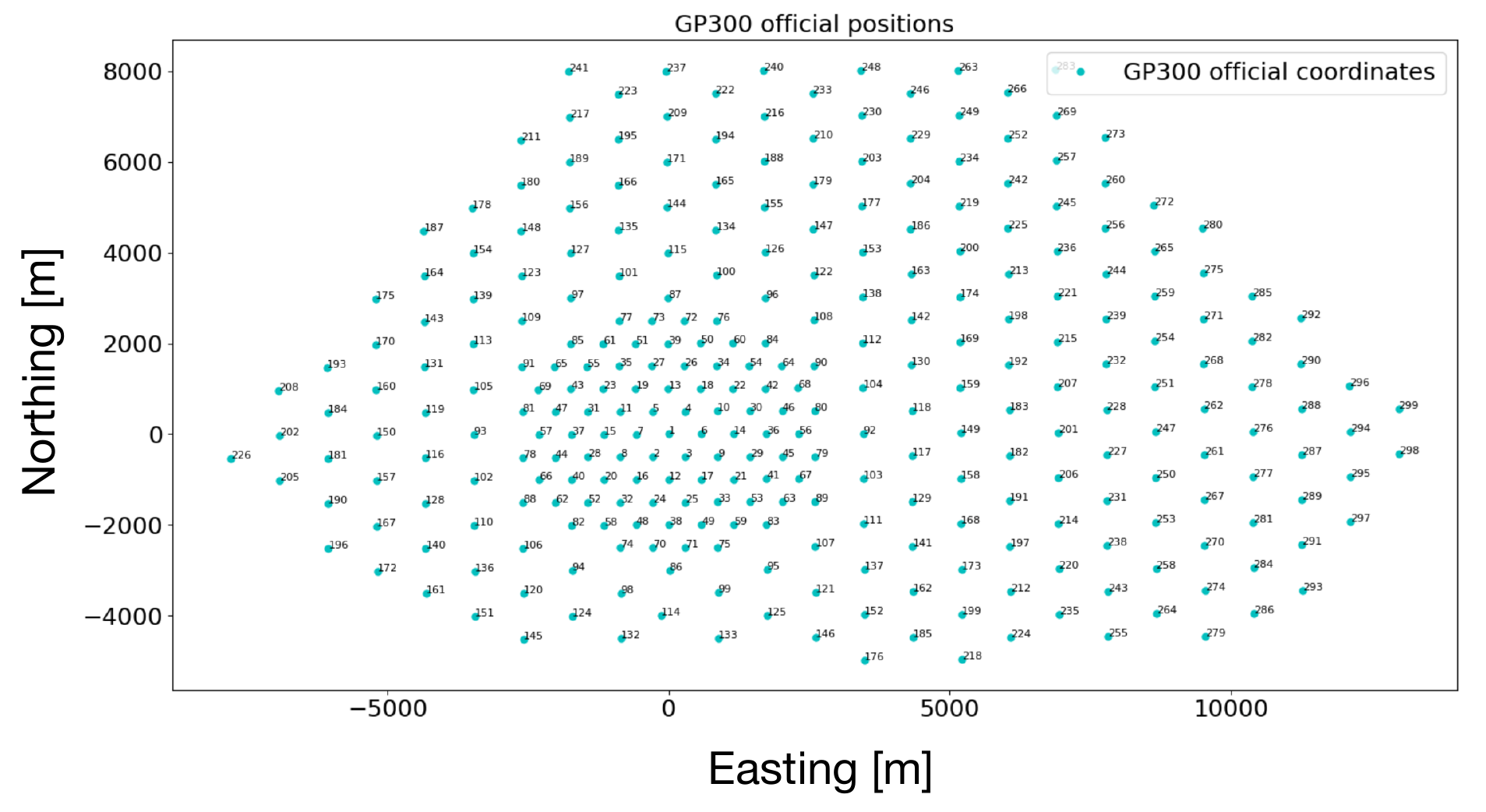}}
\caption{{\it Left:} Selected site of the GRANDProto300 radio array in Xiao Dushan (Gansu province, China).  {\it Right:} Antenna layout of GRANDProto300. A hexagonal grid was chosen with a sparse array ($1\, \rm km$-step) and a denser infill ($577\, \rm m$-step).}  
\label{fig:GP300site}
\end{figure*}

GRANDProto300 will consist of a radio array of 300 antennas deployed over $\sim 200\, \rm km^{2}$ in the radio-quiet location of Xiao Dushan (Gansu province, China). This location was chosen for its low radio background, its average altitude of $1100\, \rm m$ above sea level and its flat solid ground in a mountainous area, making it an ideal topography to detect inclined air showers. The radio array location has been officially approved by the Chinese authorities, and antenna deployment began in 2023. As shown in Fig.~\ref{fig:GP300site}, the  antenna layout consists of a hexagonal grid, combining a sparse array ($1\, \rm km$-step) with a denser infill ($577\, \rm m$-step). Thanks to its design, GRANDProto300 should be able to target the radio emission from ultra-high-energy cosmic rays and possibly gamma rays within the energy range $10^{16.5}-10^{18}\, \rm eV$, where a transition from Galactic to extragalactic sources of cosmic rays is expected to happen~\cite{Coleman}. Eventually, GRANDProto300 will act as the seed of GRAND10k~\cite{KoteraGRAND}, a radio array of $10\,000$ antennas that should be able to target the first ultra-high-energy neutrinos.

\subsection{Challenges of large-scale radio detection}
GRANDProto300 will be a test bench for the GRAND experiment. As such, it will need to address several challenges of radio detection. First, GRANDProto300 will need to reconstruct the characteristics of inclined air showers (with near-horizontal arrival directions), such as the primary particle nature, its energy, and its arrival direction. Inclined air shower detection with a sparse radio array is still an uncharted territory and is challenging because these showers undergo many effects such as ground reflection, asymmetries, or coherence loss~\cite{proc_macias_2024, Schlueter,  Chiche_2024, Guelfand_2024}, making their radio emission more complex than the one from vertical showers. This means that reconstruction algorithms accounting for these effects need to be developed and tested by GRANDProto300~\cite{proc_guelzow_2024}.

Another challenge that GRANDProto300 aims to tackle is the {\it autonomous radio detection} of high-energy astroparticles. Current experiments for air-shower radio-detection such as AERA~\cite{HuegeAERA} or LOFAR~\cite{NellesLofar}, combine radio antennas with surface detectors to detect the arrival of an air shower and then trigger the acquisition of the radio signal by the antennas. Yet, external triggers become too expensive for an experiment like GRAND with thousands of antennas. Hence, it is necessary to develop  a trigger based on the radio signals only which require to perform background rejection as early as possible in the detection chain~\cite{proc_koehler_2024, proc_correa_2024}. Towards this purpose, several promising approaches rely, for example, on the radio signal polarization~\cite{Chiche_2022}, or on machine-learning algorithms~\cite{LeCoz:2023bie}.

Eventually, GRANDProto300 will also allow us to test and validate the GRAND software~\cite{GRANDLib2024arXiv240810926G} and hardware (antenna and electronics design, reliability, etc) in experimental conditions, which will allow us to further scale the experiment to future construction stages.

 \begin{figure*}[tb]
\centering 
\includegraphics[width=0.49\columnwidth]{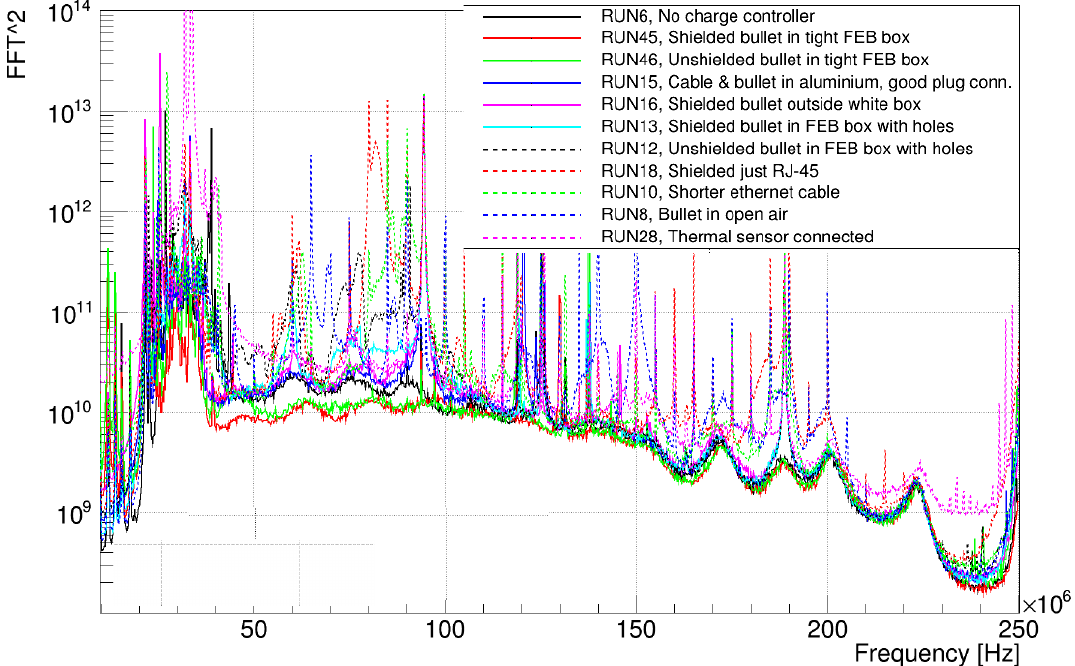}
\includegraphics[width=0.49\columnwidth]{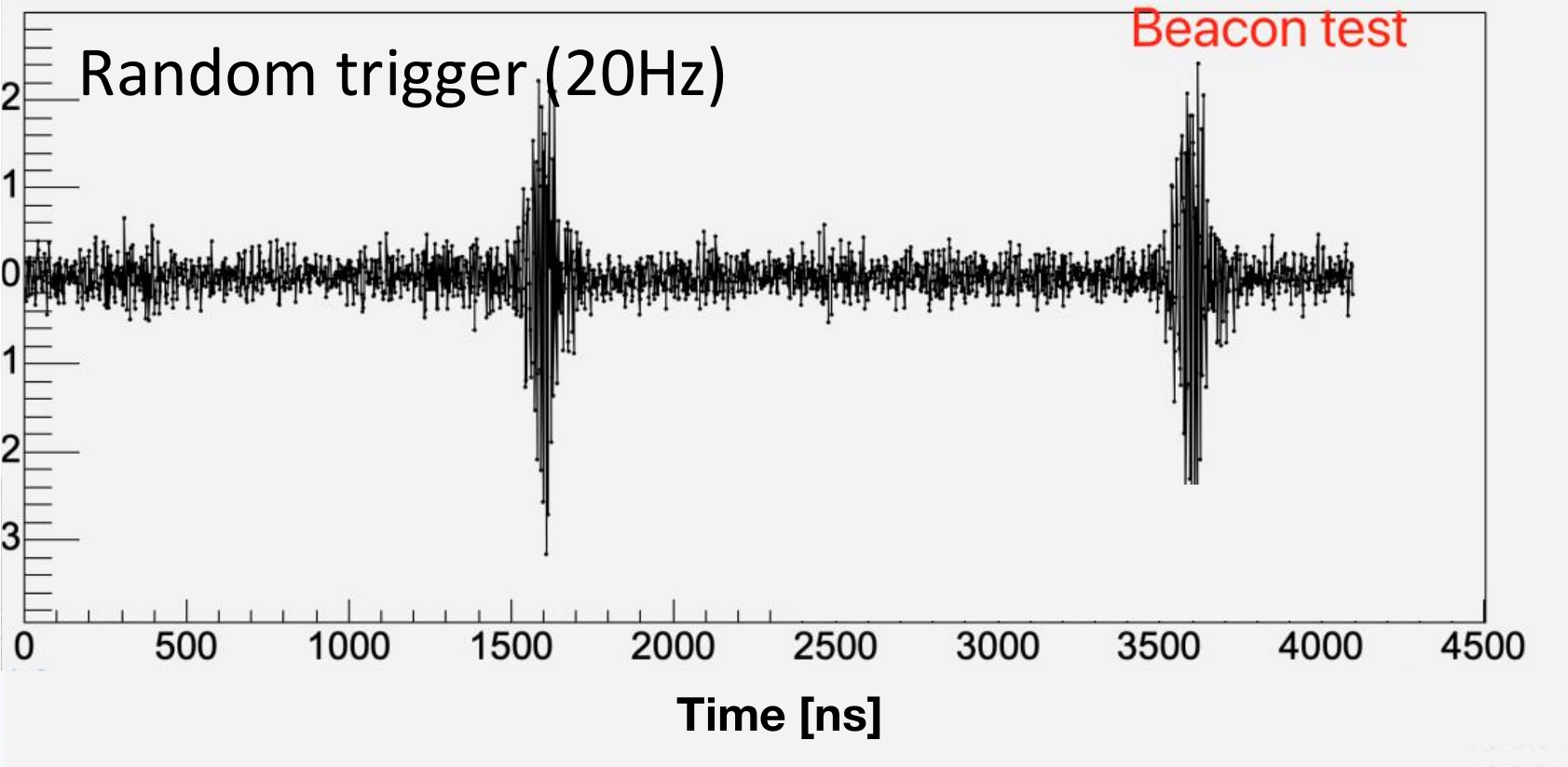}
\caption{{\it Left:} Frequency power spectra measured at the GP13 setup. For each spectrum some components are shielded/unshielded or turned on/off to test their influence on the radiations emitted by the detector.  {\it Right:}  Radio signal measured by a given GP13 triggered by a beacon fed by a sin wave at $20\, \rm Hz$ rate.
}\label{fig:Spectra}
\end{figure*}

\section{GRANDProto300 commissioning}

GRANDProto300 is currently in a commissioning phase: this means that radio antennas were deployed, the hardware was tested, and the first data are being taken.

\subsection{Detector overview}

GRANDProto300 uses butterfly antennas with three perpendicular arms to measure the three polarizations of the radio signal in the frequency range $50-200\, \rm MHz$. Each antenna is powered by solar panels, they are equipped with an low-noise amplifier (LNA) and linked to a data acquisition (DAQ) box containing a front-end board with a field-programmable gate array (FPGA) that filters the radio signal and sample it at a rate of $500\, \rm Mega \, samples/s$. Eventually, all data are transferred to a central DAQ station through a bullet Wi-Fi.
\subsection{GP13 testing and radio measurements}

 \begin{figure*}[tb]
\centering 
\includegraphics[width=0.49\columnwidth]{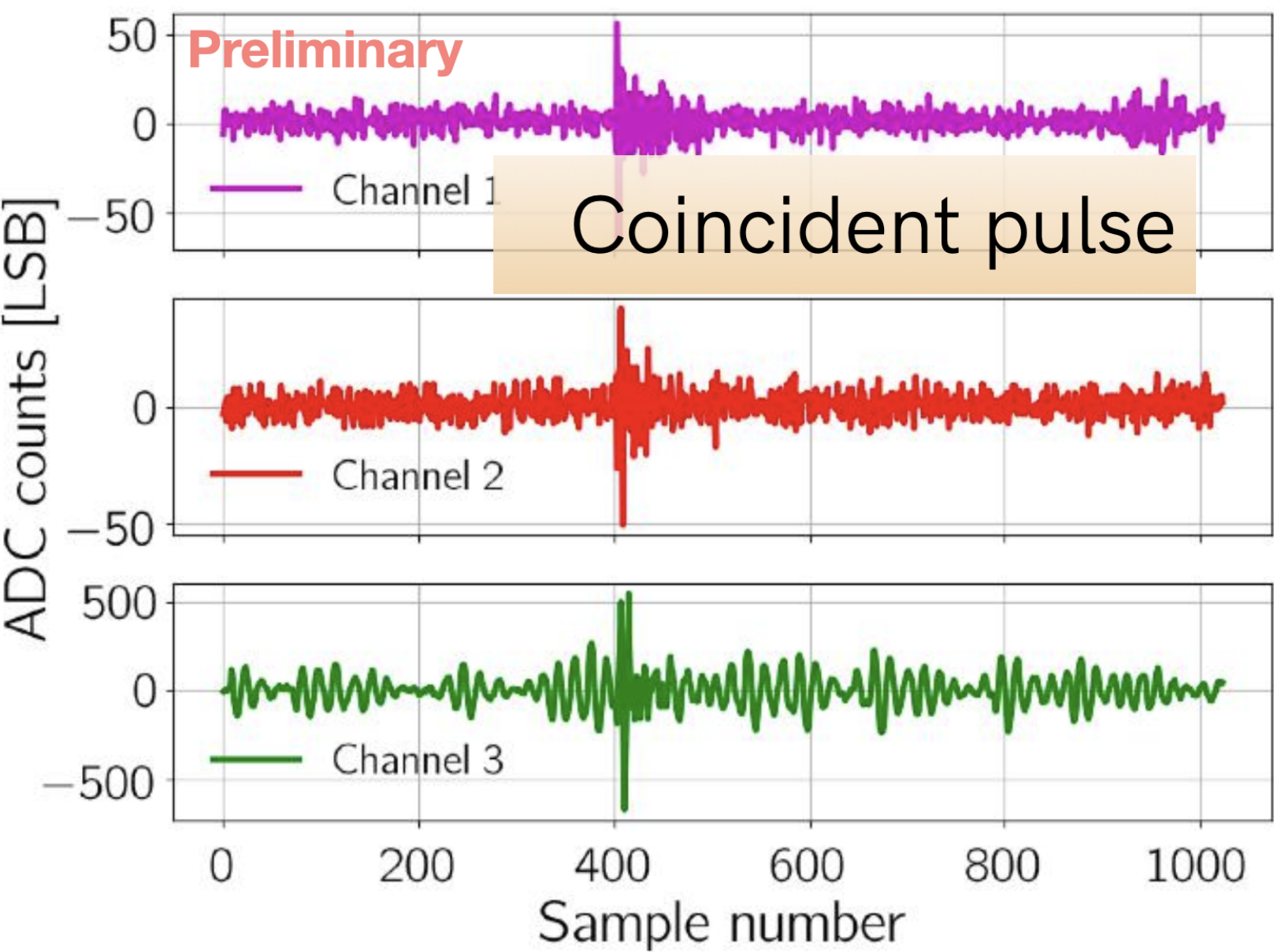}
\includegraphics[width=0.49\columnwidth]{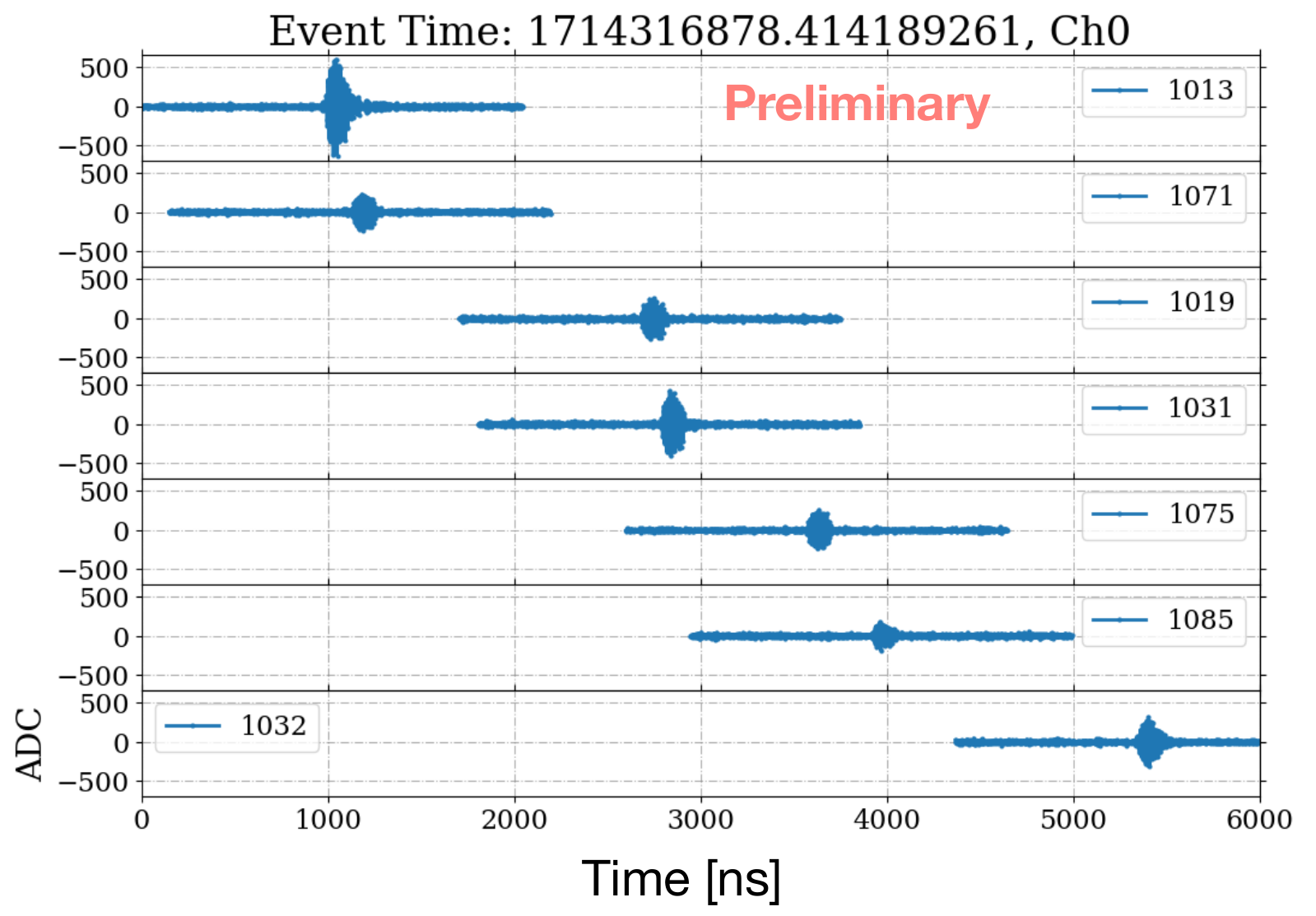}
\caption{{\it Left:} Coincident pulses observed by the different channels of a given GP13 antenna.  {\it Right:}  Time coincident radio signals observed by the y-channel (East-West) of several GP13 antennas. The delay between each time trace can be linked to the radio signal propagation.
}\label{fig:Coincidence}
\end{figure*}

The first 13 antennas were deployed in Xiao Dushan in February 2023. Using this setup, various tests of the equipment were performed. As shown in the left-hand panel of Fig.~\ref{fig:Spectra}, $\sim 40$ data measurements of frequency spectra were realized to test the influence of each parameter. The conclusion was that all components of the detection unit radiate. Additionally, the tests revealed a heating issue with the LNA and the front-end board. To solve this issue, all the components have been shielded and the heating issue was fixed, which allowed for radio measurements with much cleaner frequency spectra. As shown in the right-hand panel of Fig.~\ref{fig:Spectra}, GP13 then successfully triggered on the signal from a beacon antenna fed with a sine wave. 

After testing the equipment, the first data were taken.  As shown in Fig.~\ref{fig:Coincidence}, coincident pulses were observed between the different channels of a given antenna (left-hand panel) and radio signals were observed with temporal coincidence between several antennas (right-hand panel). 

\begin{figure}[htbp]
     \vspace{-0.4cm}

    \centering
    \begin{minipage}[b]{0.55\textwidth}
        \centering
        \includegraphics[width=\textwidth]{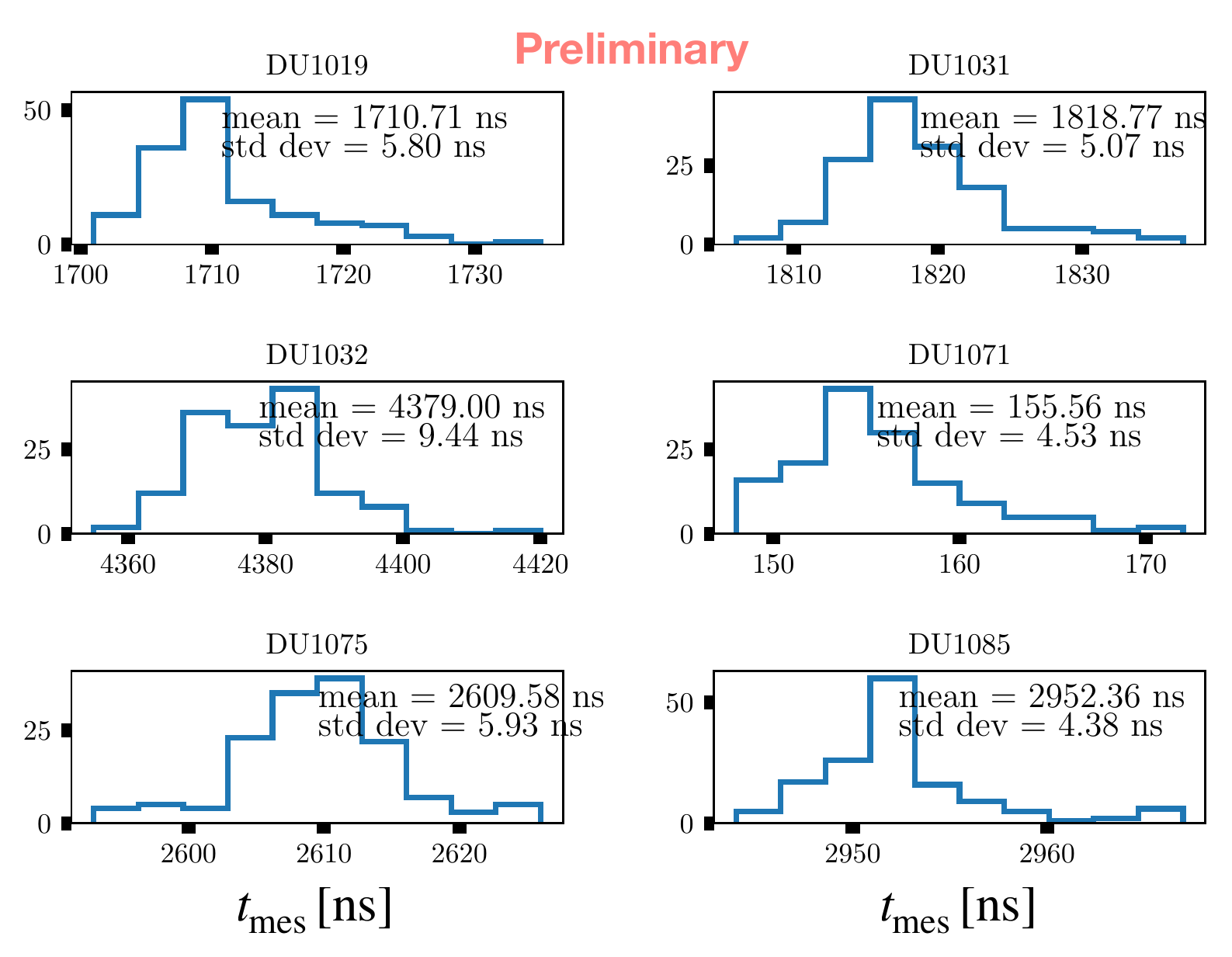}
    \end{minipage}
    \hfill
    \begin{minipage}[b]{0.44\textwidth}
        \centering
        \raisebox{0.4cm}{\includegraphics[width=\textwidth]{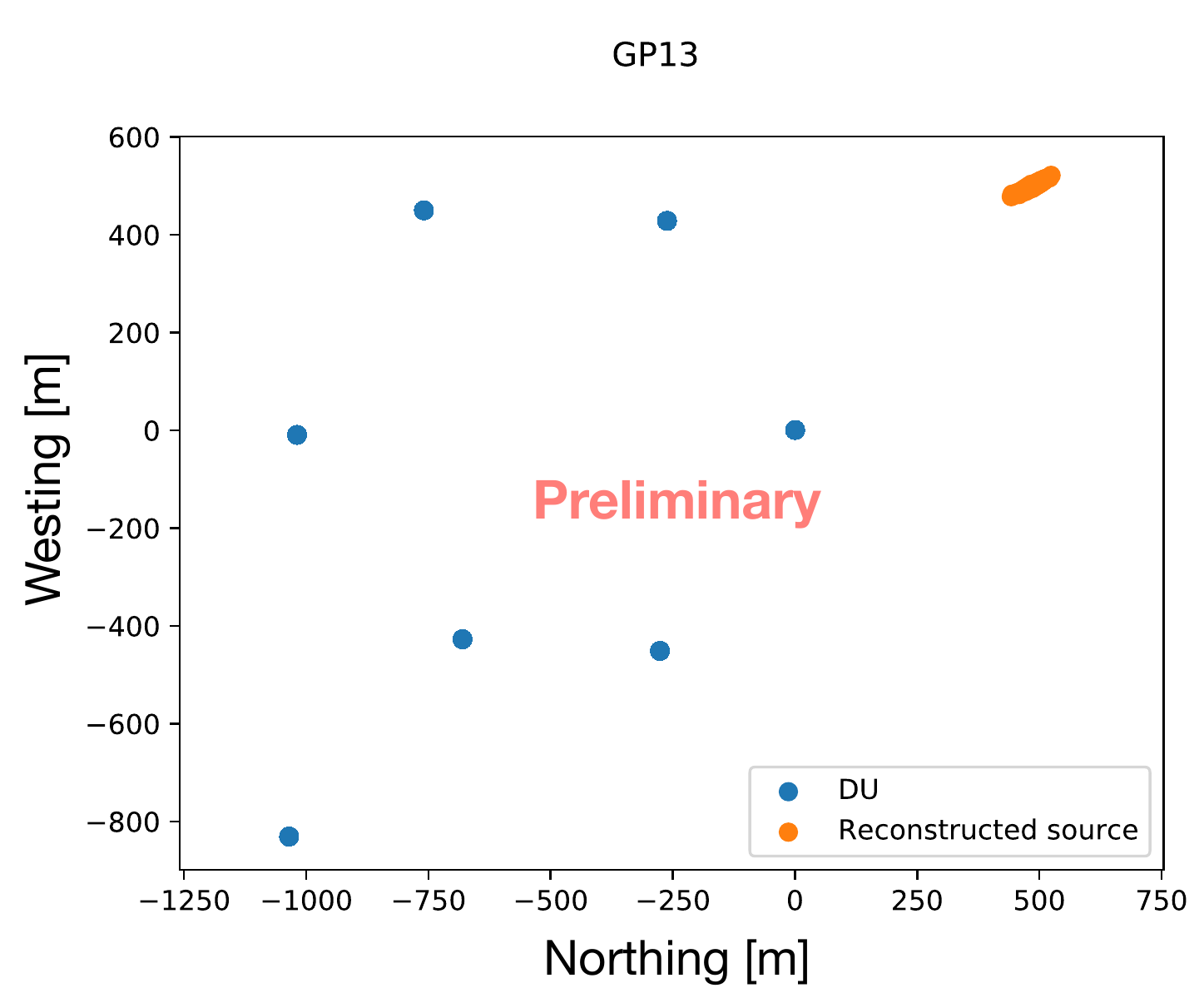}}
    \end{minipage}


    \begin{minipage}[b]{0.38\textwidth}
        \centering
        \includegraphics[width=\textwidth]{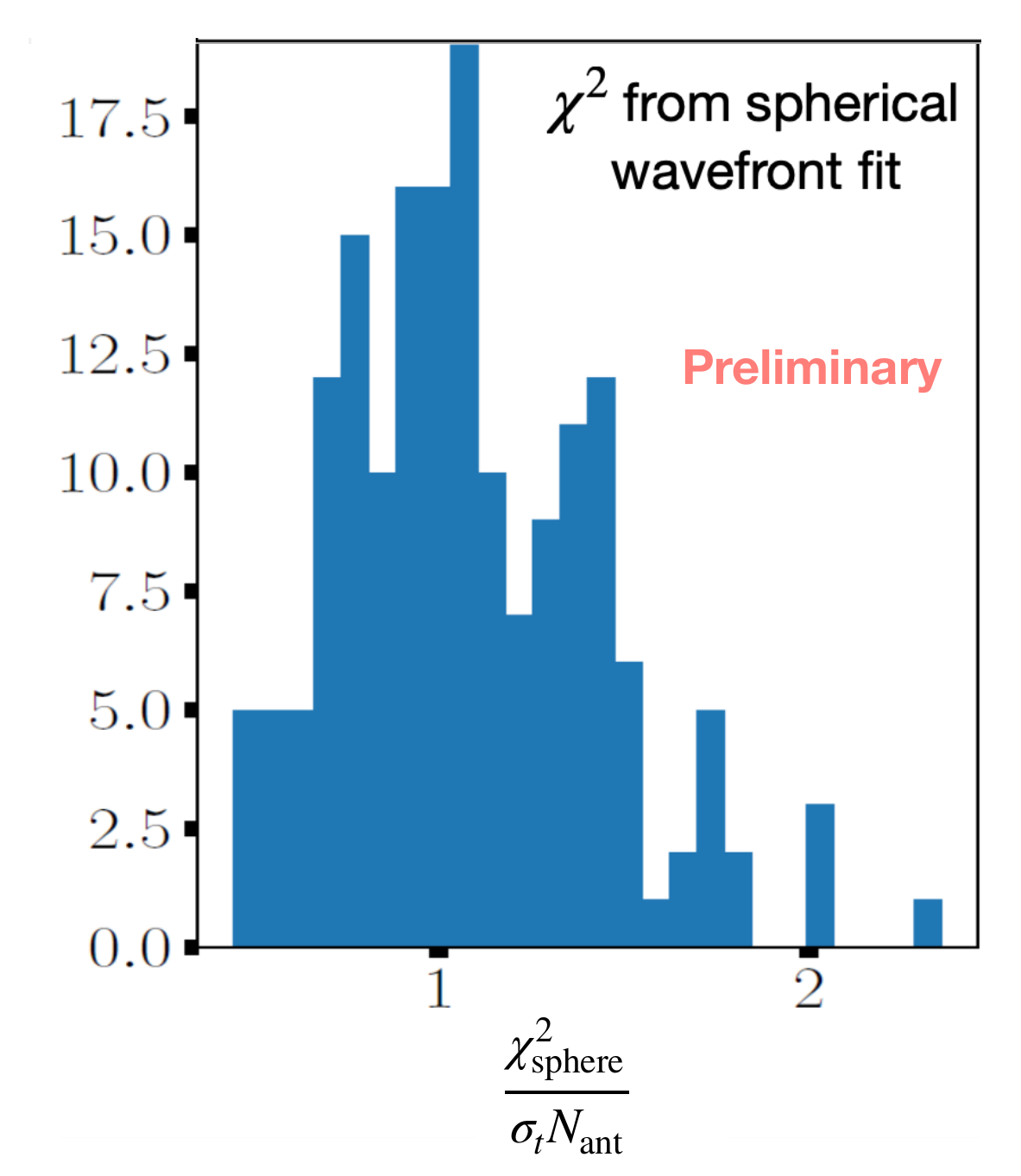}
    \end{minipage}
    \hfill
    \begin{minipage}[b]{0.6\textwidth}
        \centering
        \raisebox{0.1cm}{\includegraphics[width=\textwidth]{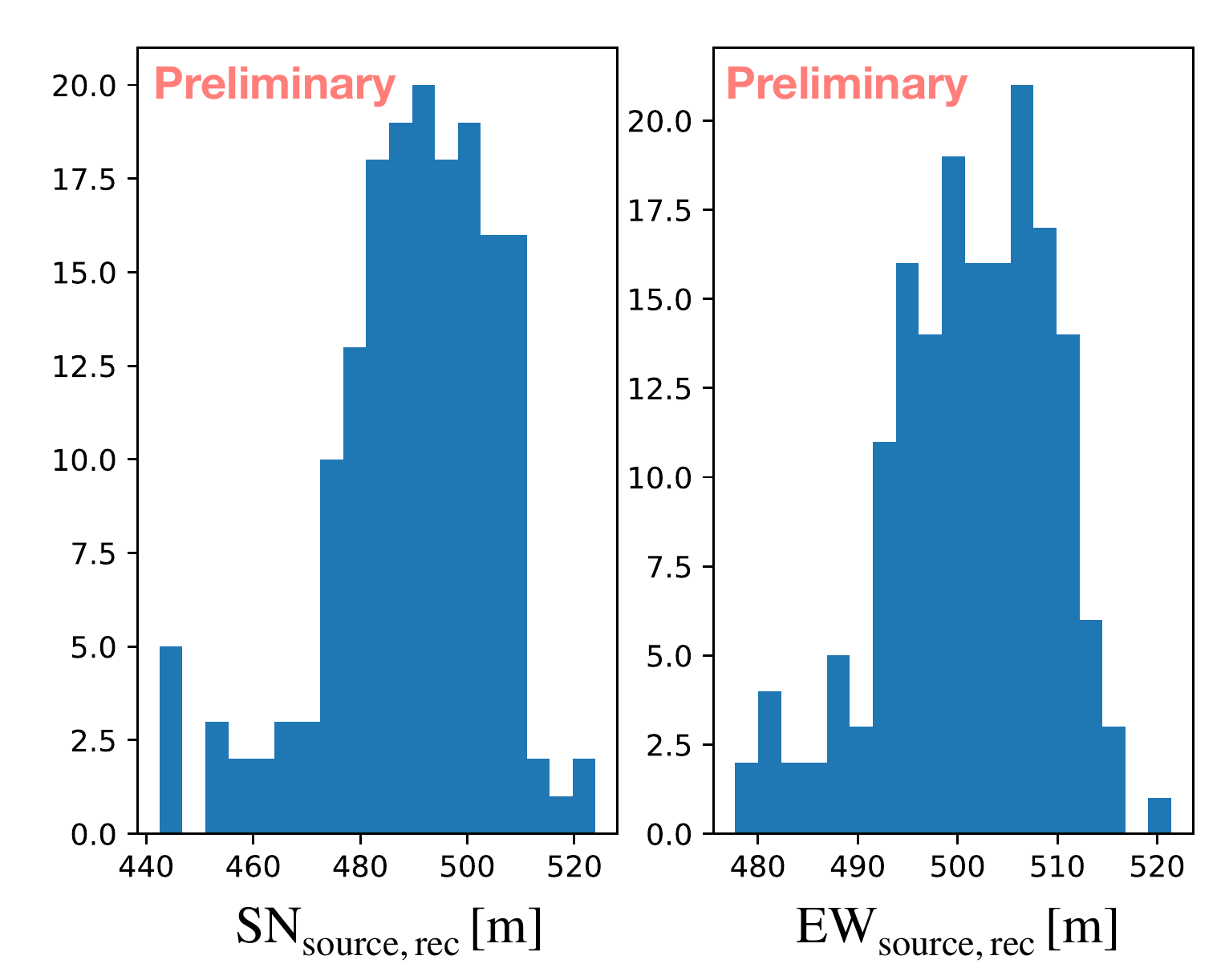}}
    \end{minipage}
    
    \caption{Reconstruction of a beacon antenna position. {\it Top-left:} Distribution of trigger times at several detection units. {\it Bottom-left:} Normalized $\chi^{2}$ from a spherical wavefront fit. {\it Top-right:} Reconstructed position in the ground plane. {\it Bottom-right:} Distributions of reconstructed Easting and Northing.}\label{fig:BeaconRec}
\end{figure}

Eventually, using the timing information at the antenna level, the position of a beacon antenna located on top of the central DAQ station was reconstructed, as shown in Fig.~\ref{fig:BeaconRec}. The top-left panel shows the distribution of the trigger times at several detection units. The results yield a standard deviation of the trigger times of $\sim 5\, \rm ns$ for most units, which meets the expectations targeted by GRAND for an accurate reconstruction. The timing information was then used to evaluate a normalized $\chi^{2}$, i.e., the summed squared differences between the measured data and a model predicting the trigger times, assuming a spherical wavefront for the radio signal (bottom-left panel). Quality cuts are applied to keep only signals with a normalized $\chi^{2}<5$. All the events that pass this cut (171 out of 172 in this case) were then used to reconstruct the beacon position, The results are shown on the top-right and bottom-right panels. The position of the beacon was successfully reconstructed with a standard deviation of $\sim 15\, \rm m$ on the Northing and of $\sim 8\, \rm m$ on the Easting.

\subsection{Next stages of GRANDProto300}

The GP13 radio array will soon be expanded to GP80, with 70 additional antennas to be deployed by the end of 2024. By then, the firmware needs to be updated, an improved communication between the detection units and the central DAQ is needed and the aim is to reconstruct the arrival direction of other sources. With 83 antennas (current 13 and 70 additional, evenly spaced), triggered rate predictions estimate that, with a conservative threshold, GP80 should be able to detect $\sim 30$ cosmic-ray induced events per day, with primary energies between $2\times 10^{17}- 2\times 10^{18}\, \rm eV$. The main objective of GP80 will therefore be to reconstruct the first cosmic-ray events.

The GRANDProto300 stage is then expected by $\sim 2026$, with $\mathcal{O}(300)$ radio antennas over $200\, \rm km^{2}$, and could also be completed with surface detectors to help calibrate the detector. This radio array should be able to detect ultra-high-energy cosmic-rays and possibly ultra-high-energy gamma rays (if complemented by surface detectors) in the energy range $10^{16.5}-10^{18}\, \rm eV$ and will validate the GRAND detection principle allowing the experiment to scale up to the further stages.

\section{Expected performances}
GRANDProto300 will be the first experiment to detect inclined air showers using a sparse radio array. An accurate reconstruction of air showers will therefore be crucial to validate the GRAND detection principle. We discuss below the expected performances of GRANDProto300.
\subsection{Antenna layout and trigger rate}

\begin{figure*}[tb]
\centering 
\includegraphics[width=0.49\columnwidth]{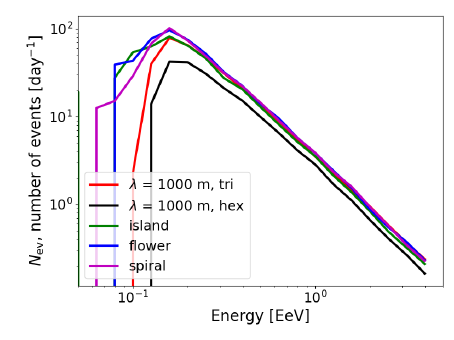}
\includegraphics[width=0.49\columnwidth]{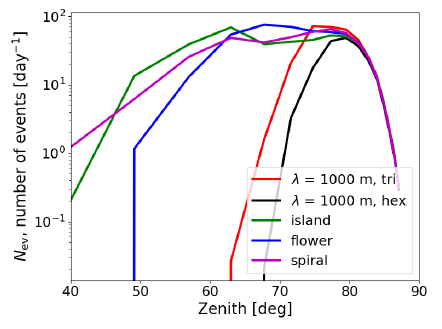}
\caption{Trigger rates for different antenna geometries as a function of the primary particle energy ({\it left}) and zenith angle ({\it right}). The `tri' and `hex' geometries correspond to a triangular and hexagonal grid respectively and have no infill. The `hex' geometry is for 150 antennas, while all other geometries are for $\sim 250$ antennas. The `island', `flower' and `spiral' correspond to hexagonal girds with an infill (from~\cite{BenoitL2024JInst..19P4006B}).
}\label{fig:TriggerRate}
\end{figure*}

Different antenna configurations were tested to evaluate and optimize the GRANDProto300 trigger rate using ZHAireS Monte-Carlo simulations~\cite{ZHS_2012APh....35..325A, BenoitL2024JInst..19P4006B}. The study found that a hexagonal grid of antennas provides the best performance.  Additionally, the effect of adding a dense infill on top of the hexagonal grid was tested, considering different geometries and a sparse array with a $1\, \rm km$-spacing. The trigger rate was estimated by fixing a trigger threshold at the antenna level of $V=75\, \rm \mu V$ and requiring that at least $N_{\rm trig}=5$ neighboring antennas are triggered to consider an event triggered. The trigger rate for different geometries is shown in Fig.~\ref{fig:TriggerRate} as a function of the primary particle energy (left-hand panel) and zenith angle (right-hand panel). Both plots demonstrate that the hexagonal and triangular geometries yield similar trigger rates, despite the triangular configuration having approximately 100 more antennas. Moreover, the results indicate that the inclusion of an infill enhances detection efficiency.

\subsection{$X_{\rm max}$ and angular resolution}

\begin{figure*}[tb]
\centering 
\includegraphics[width=0.49\columnwidth]{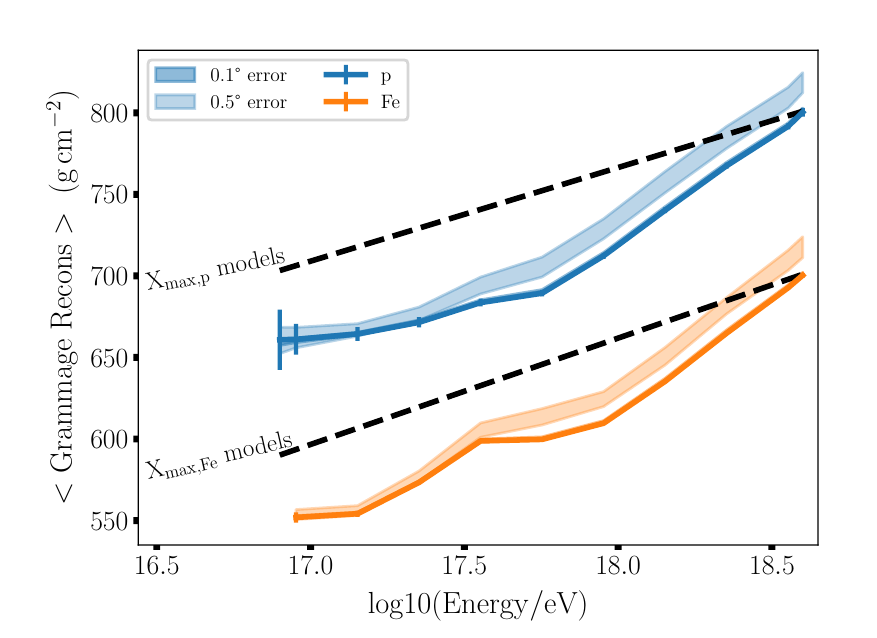}
\includegraphics[width=0.49\columnwidth]{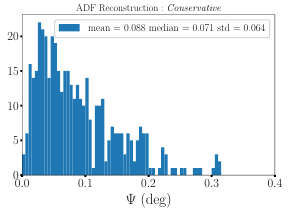}
\caption{({\it Left}) $X_{\rm max}$ reconstruction expected with GRANDProto300 for proton (blue) and iron (orange) primaries. The thick line assumes an error on the angular reconstruction of $0.1^{\circ}$, while the shaded area corresponds to an error of $0.5^{\circ}$  (from~\cite{DecoeneXmax2021arXiv211207542D}). ({\it Right}) Angular resolution estimated with ZHAireS simulations for a GRANProto300-like layout (from~\cite{DecoenePhD}).
}\label{fig:Rec}
 \vspace{-0.3cm}
\end{figure*}

The reconstruction of the depth of shower maximum, $X_{\rm max}$, and of the shower direction were also evaluated with ZHAireS simulations. The electric field amplitude at the antenna level was first described with a phenomenological model that exploits the symmetries of the radio signal to derive an angular distribution function (ADF)~\cite{DecoeneADF2022icrc.confE.211D, proc_macias_2024}. The ADF depends on the shower arrival direction and on the radio signal emission point, which are treated as free parameters.  Assuming a spherical wavefront model, these parameters were fitted using a $\chi^{2}$ minimization  based on the measured amplitudes at the antenna level~\cite{DecoenePhD, DecoeneADF2022icrc.confE.211D}. On the left-hand panel of Fig.~\ref{fig:Rec} we show the reconstructed $X_{\rm max}$ as a function of the primary particle energy. The results show a similar behaviour between the reconstructed $X_{\rm max}$ and the models. Additionally, the standard deviation on $X_{\rm max} $ was shown to be comparable to that achieved by current radio experiments~\cite{DecoeneXmax2021arXiv211207542D}. On the right-hand panel of Fig.~\ref{fig:Rec} we show that with a GRANDProto300-like layout we reach a mean and median angular resolution below $0.1^{\circ}$. If confirmed experimentally, this performance will be one of the main assets of the GRAND experiment, making possible the identification of the first ultra-high-energy point sources and opening the path toward an ultra-high-energy neutrino astronomy.
\section{Science case}
Thanks to its design and expected performance, GRANDProto300 will have a broad science case covering astroparticle physics and radio astronomy.
\subsection{Galactic-to-extragalactic transition}
GRANDProto300 will be able to detect cosmic rays in the energy range $10^{16.5}-10^{18}\, \rm eV$, between the knee and the second knee of the cosmic-ray spectrum, where a Galactic-to-extragalactic transition of sources is believed to happen~\cite{Coleman}. GRANDProto300 will  add more statistics in this transition region and provide the first independent measurement of the cosmic-ray spectrum based on the radio signal only. 
Additionally, if complemented with particle detectors, GRANDProto300 will be able to measure the shower electromagnetic and muonic content independently, providing one of the most efficient ways to reconstruct the primary particle nature and to infer an event-by-event mass composition~\cite{Holt}.
 \vspace{-0.1cm}
\subsection{Ultra-high-energy gamma rays}
Ultra-high-energy gamma rays ($>10^{17}\,\rm eV$) are expected to be produced by the interaction of ultra-high-energy cosmic rays with cosmological photon backgrounds~\cite{Zatsepin:1966jv}. Yet, their detection is challenging since their interaction cross section limits their horizon to $\sim 1\, \rm Mpc$. Since almost no muons are expected in gamma-ray-induced air showers, an independent measurement of the shower electromagnetic and muonic content would provide an efficient way to identify gamma ray primaries. Hence, if GRANDProto300 is complemented by particle detectors, it would possibly detect the first ultra-high-energy gamma rays or at least put new limits on their flux.
\subsection{Fast radio bursts}
Eventually, GRANDProto300 should be able to detect fast radio bursts (FRB), powerful transient radio pulses with a typical duration of a few ms~\cite{BzhangFRB2023RvMP...95c5005Z}. Thanks to its large field of view and high sensitivity, GRANDProto300 is well-suited to perform FRB searches. Investigations are in progress to evaluate whether GRANDProto300 should target FRBs using the unphased radio signal or beam-forming. In the latter case, we estimate that GRANDProto300 should detect $\sim 1$ FRB per month.

\section{Conclusion}

GRANDProto300 is a 300-antennas radio array that will detect cosmic rays in the energy range $10^{16.5}-10^{18}\, \rm eV$. The first 13 antennas were deployed in 2023 and preliminary data are encouraging. The next step will be the deployment of 70 additional antennas by 2025, which should allow for the first detection of a cosmic ray event. The full GRANDProto300 radio array is expected by 2026 and will cover several physics cases, including the Galactic-to-extragalctic transition, ultra-high-energy gamma rays, and fast radio bursts. GRANDProto300 will then be expanded to GRAND10k in the 2030s, which will target the detection of the first ultra-high-energy neutrinos. 

\bibliographystyle{elsarticle-num}
{\footnotesize
\bibliography{references}

\begin{thebibliography}{10}
\expandafter\ifx\csname url\endcsname\relax
  \def\url#1{\texttt{#1}}\fi
\expandafter\ifx\csname urlprefix\endcsname\relax\def\urlprefix{URL }\fi
\expandafter\ifx\csname href\endcsname\relax
  \def\href#1#2{#2} \def\path#1{#1}\fi

\bibitem{GRAND}
{GRAND Collaboration}, Science China Physics, Mechanics, and Astronomy 63~(1) (2020) 219501.

\bibitem{KoteraGRAND}
K.~{Kotera}, PoS ARENA2024 (2024) 057.
\newblock \href {https://doi.org/10.48550/arXiv.2408.16316} {\path{doi:10.48550/arXiv.2408.16316}}.

\bibitem{Coleman}
A.~{Coleman}, et~al., Astroparticle Physics 147 (2023) 102794.
\newblock \href {https://doi.org/10.1016/j.astropartphys.2022.102794} {\path{doi:10.1016/j.astropartphys.2022.102794}}.

\bibitem{proc_macias_2024}
O.~Macias, PoS ARENA2024 (2024) 057.
\newblock \href {https://doi.org/10.48550/arXiv.2408.15952} {\path{doi:10.48550/arXiv.2408.15952}}.

\bibitem{Schlueter}
F.~{Schl{\"u}ter}, et~al., European Physical Journal C 80~(7) (2020) 643.
\newblock \href {https://doi.org/10.1140/epjc/s10052-020-8216-z} {\path{doi:10.1140/epjc/s10052-020-8216-z}}.

\bibitem{Chiche_2024}
S.~{Chiche}, et~al., Physical Review Letters 132~(23) (2024) 231001.
\newblock \href {https://doi.org/10.1103/PhysRevLett.132.231001} {\path{doi:10.1103/PhysRevLett.132.231001}}.

\bibitem{Guelfand_2024}
M.~{Guelfand}, et~al., JCAP 2024~(5) (2024) 055.
\newblock \href {https://doi.org/10.1088/1475-7516/2024/05/055} {\path{doi:10.1088/1475-7516/2024/05/055}}.

\bibitem{proc_guelzow_2024}
L.~G{\"u}lzow, PoS ARENA2024 (2024) 057.

\bibitem{HuegeAERA}
T.~{Huege}, {Pierre Auger Collaboration}, in: The European Physical Journal C, Vol. 210, 2019, p. 05011.

\bibitem{NellesLofar}
A.~{Nelles}, et~al., JCAP 2015~(5) (2015) 018--018.
\newblock \href {https://doi.org/10.1088/1475-7516/2015/05/018} {\path{doi:10.1088/1475-7516/2015/05/018}}.

\bibitem{proc_koehler_2024}
J.~K{\"o}hler, PoS ARENA2024 (2024) 061.

\bibitem{proc_correa_2024}
P.~Correa, PoS ARENA2024 (2024) 057.

\bibitem{Chiche_2022}
S.~{Chiche}, et~al., Astroparticle Physics 139 (2022) 102696.

\bibitem{LeCoz:2023bie}
S.~Le~Coz, PoS ICRC2023 (2023) 224.
\newblock \href {https://doi.org/10.22323/1.444.0224} {\path{doi:10.22323/1.444.0224}}.

\bibitem{GRANDLib2024arXiv240810926G}
{GRAND Collaboration}, arXiv e-prints (Aug. 2024).
\newblock \href {https://doi.org/10.48550/arXiv.2408.10926} {\path{doi:10.48550/arXiv.2408.10926}}.

\bibitem{BenoitL2024JInst..19P4006B}
A.~{Benoit-L{\'e}vy}, et~al., JINST 19~(4) (2024) P04006.
\newblock \href {https://doi.org/10.1088/1748-0221/19/04/P04006} {\path{doi:10.1088/1748-0221/19/04/P04006}}.

\bibitem{ZHS_2012APh....35..325A}
J.~{Alvarez-Mu{\~n}iz}, et~al., Astropart. Phys. 35~(6) (2012) 325--341.

\bibitem{DecoeneXmax2021arXiv211207542D}
V.~Decoene, et~al., Astropart. Phys. 145 (2023) 102779.
\newblock \href {https://doi.org/10.1016/j.astropartphys.2022.102779} {\path{doi:10.1016/j.astropartphys.2022.102779}}.

\bibitem{DecoenePhD}
V.~Decoene, Theses, {Sorbonne Universit{\'e}} (2020).

\bibitem{DecoeneADF2022icrc.confE.211D}
V.~{Decoene}, et~al., in: 37th International Cosmic Ray Conference, 2022, p. 211.
\newblock \href {https://doi.org/10.22323/1.395.0211} {\path{doi:10.22323/1.395.0211}}.

\bibitem{Holt}
E.~Holt, et~al., The European Physical Journal C 79 (2019) 371.
\newblock \href {https://doi.org/10.1140/epjc/s10052-019-6859-4} {\path{doi:10.1140/epjc/s10052-019-6859-4}}.

\bibitem{Zatsepin:1966jv}
G.~T. Zatsepin, V.~A. Kuzmin, JETP Lett. 4 (1966) 78--80.

\bibitem{BzhangFRB2023RvMP...95c5005Z}
B.~{Zhang}, Reviews of Modern Physics 95~(3) (2023) 035005.
\newblock \href {https://doi.org/10.1103/RevModPhys.95.035005} {\path{doi:10.1103/RevModPhys.95.035005}}.

\end{thebibliography}
}

\end{document}